# Coordinated Analysis of Two Graphite Grains from the CO3.0 LAP 031117 Meteorite: First Identification of a CO Nova Graphite and a Presolar Iron Sulfide Subgrain

Pierre Haenecour[1,2,3,4*], Christine Floss[1,3,4], Jordi José [5], Sachiko Amari[1,3,4], Katharina Lodders[2,4], Manavi Jadhav[6], Alian Wang[2,4], and Frank Gyngard[1,3,4]

[1]Laboratory for Space Sciences, Washington University in St. Louis, One Brookings Drive, St. Louis, MO 63130-4899, USA.

[2]Department of Earth and Planetary Sciences, Washington University in St. Louis, One Brookings Drive, St. Louis, MO 63130-4899, USA.

[3]Physics Department, Washington University in St. Louis, One Brookings Drive, St. Louis, MO 63130-4899, USA.

[4]McDonnell Center for the Space Sciences, Washington University in St. Louis, One Brookings Drive, St. Louis, MO 63130-4899, USA.

[5]Departament de Física, EUETIB, C. Urgell 187, Universitat Politècnica de Catalunya, E-08036 Barcelona, Spain; Institut d'Estudis Espacials de Catalunya, Ed. Nexus-21, C. Gran Capità 2-4,E-08034 Barcelona, Spain

[6]Department of Physics, University of Louisiana at Lafayette, Broussard Hall, P.O. Box 43680, Lafayette, LA 70504-3680, USA.

[*]Corresponding Author: P. Haenecour. Email: haenecour@wustl.edu. Phone: (314) 935-6243. Address: One Brookings Drive, Campus Box 1169, St. Louis, MO 63130, USA.

# ABSTRACT

Presolar grains constitute remnants of stars that existed before the formation of the solar system. In addition to providing direct information on the materials from which the solar system formed, these grains provide ground-truth information for models of stellar evolution and nucleosynthesis. Here we report the *in-situ* identification of two unique presolar graphite grains from the primitive meteorite LaPaz Icefield 031117. Based on these two graphite grains, we estimate a bulk presolar graphite abundance of $5^{+7}_{-3}$ ppm in this meteorite. One of the grains (LAP-141) is characterized by an enrichment in $^{12}$C and depletions in $^{33,34}$S, and contains a small iron sulfide subgrain, representing the first unambiguous identification of presolar iron sulfide. The other grain (LAP-149) is extremely $^{13}$C-rich and $^{15}$N-poor, with one of the lowest $^{12}$C/$^{13}$C ratios observed among presolar grains. Comparison of its isotopic compositions with new stellar nucleosynthesis and dust condensation models indicates an origin in the ejecta of a low-mass CO nova. Grain LAP-149 is the first putative nova grain that quantitatively best matches nova model predictions, providing the first strong evidence for graphite condensation in nova ejecta. Our discovery confirms that CO nova graphite and presolar iron sulfide contributed to the original building blocks of the solar system.

# 1. INTRODUCTION

About 4.6 billion years ago, our solar system formed from a molecular cloud composed of gas, ice, and dust. It was initially believed that, during the early stages of solar system formation, all of the original circumstellar and interstellar dust grains were vaporized, leaving an isotopically homogenized solar nebula (Cameron, 1962). However, this theory was challenged by the identification of noble gas isotopic anomalies in meteorites, providing the first hint for the survival of circumstellar grains (also called 'presolar grains') (Black & Pepin, 1969; Reynolds & Turner, 1964). Circumstellar grains can be identified by their highly anomalous isotopic compositions in major (e.g., C, O) and minor elements compared to the average isotopic composition of solar system materials (Zinner, 2014). Characterization of the isotopic and elemental compositions of presolar grains opened a new field in astronomy and astrophysics, allowing the direct study of individual stars (Zinner, 2014) and providing ground-truth information on stellar evolution, nucleosynthesis of the elements, and grain condensation in circumstellar envelopes. As presolar grains are the original building blocks of the solar system, the *in situ* survey of fine-grained material in primitive meteorites has also allowed the identification of new presolar grain types, such as wüstite, silica and magnetite (Floss et al. 2008; Haenecour et al. 2013; Zega et al. 2015). Indeed, presolar grains consist of a variety of minerals and/or amorphous assemblages, including carbonaceous grains (e.g., diamond, graphite, silicon carbide), oxides (e.g. corundum, spinel, hibonite), and silicates (e.g., olivine, pyroxene, silica) (Floss & Haenecour, 2016; Zinner, 2014). Presolar grains can also contain a variety of inclusions (e.g., metal nuggets, carbides, oxides) (Croat et al. 2014; Zinner, 2014).

As the carrier phase of a $^{22}$Ne-rich component (termed Ne-E(L)), presolar graphites were first isolated through complex chemical and physical methods, including density separations and harsh chemical-dissolution treatments to dissolve the bulk of the meteorite material and concentrate the acid-resistant phases into residues (Amari et al. 1990, 1994). Over the past 25 years, more than 2000 presolar graphite grains have been studied from only two meteorites, Murchison (CM2) and Orgueil (CI1) (Amari et al. 2014; Hoppe et al. 1995; Jadhav et al. 2013b; Zinner et al. 1995). They range in density between 1.6–2.2 **g/cm$^3$** and are often classified into two groups: low-density (LD) and high-density (HD) graphites. They also exhibit two main morphologies (cauliflower aggregates and shell-like onions), each corresponding to different degrees of graphitization (from poorly graphitic to well-crystallized) (Croat et al. 2014; Zinner, 2014). High-precision multi-element isotopic studies indicate that most presolar graphite grains originate from low-metallicity asymptotic giant branch (AGB) stars (~50%) and supernovae (~25%) (Amari et al. 2014; Jadhav et al. 2013a, 2014); a small fraction might have also originated from born-again AGB stars and J-type stars (Jadhav et al. 2013b).

Here, we report the first *in situ* identification of two presolar graphite grains in a primitive meteorite. The grains (LAP-141 and LAP-149) were identified by NanoSIMS ion imaging in a thin section of the primitive meteorite LAP 031117.

## 2. SAMPLE AND EXPERIMENTAL METHODS

A petrographic thin-section of the type 3.00 CO carbonaceous chondrite LaPaz Icefield 031117 (LAP 031117) was obtained from the NASA Johnson Space Center meteorite curatorial facility. Based on the distribution of $Cr_2O_3$ in ferroan olivine from chondrules and the fayalite contents of amoeboid olivine inclusions (AOIs), Chizmadia & Cabret-Lebron (2009) showed that LAP 031117 is very primitive, much like the CO3.00 chondrite ALHA77307. Apart from the presence of minor sporadic terrestrial alteration (Fe-rich) veins, they did not find any sign of thermal metamorphism and/or aqueous alteration as commonly found in CO3 of higher petrologic subtypes (> CO3.1). LAP 031117 is also characterized by presolar grain abundances similar to the most primitive meteorites (P. Haenecour et al. 2016, in preparation). A transmission electron microscopy study of several distinct areas in LAP 031117 does show the presence of localized (micrometer-scale) aqueous alteration in a fine-grained rim around a chondrule (P. Haenecour et al. 2016, in preparation); however, no signs of alteration were found in the area where the two graphite grains were identified.

The grains were identified by NanoSIMS 50 raster ion imaging during a search for O- and C-anomalous presolar grains. A focused $Cs^+$ primary beam of ~1 pA, with a diameter of about 100 nm was rastered over fine-grained matrix and rims around chondrules, and secondary ions of $^{12,13}C^-$ and $^{16,17,18}O^-$, as well as secondary electrons (SE), were simultaneously acquired in multicollection mode. Each measurement consisted of 5-10 scans of $10 \times 10$ μm$^2$ ($256 \times 256$ pixels) areas rastered within $12 \times 12$ μm$^2$ regions pre-sputtered to remove the carbon coat. The individual imaging layers were sequentially added to produce a single cumulative image for each isotope analyzed. Isotope ratio images were calculated from the ion images and used to identify isotopically anomalous grains. A grain was considered presolar if its isotopic composition deviated from that of the average surrounding material by more than 5σ, and the anomaly was present in at least three consecutive layers. Using similar analytical conditions, we also acquired nitrogen, silicon and sulfur isotopic maps of grain LAP-149 in two separate measurement sessions: $^{12}C^{14}N^-$, $^{12}C^{15}N^-$, $^{28,29,30}Si^-$ in one session, and $^{12,13}C^-$, $^{32,33,34}S^-$ in the other; the sulfur isotopic composition of LAP-141 was also measured during this second session. For sulfur isotopic measurements, the Quasi-Simultaneous Arrival (QSA) effect is expected to introduce a non-negligible bias in isotopic measurements (Slodzian et al. 2004). We estimated the correction factor for the QSA effect by measuring the sulfur isotopic composition of the surrounding matrix area, as well as a troilite grain in LAP 031117, and changing the secondary ion transmission by adjusting the aperture slit width, as has been done previously (Nishizawa et al. 2010; Slodzian et al. 2004). However, we did not find any evidence for the QSA effect and thus did not apply a correction factor. We also regularly measured the S isotopic composition of surrounding matrix material (in several areas) to ensure reproducibility. The sulfur isotopic compositions are normalized to the sulfur isotopic composition of the surrounding matrix, expressed in δS-values as permil deviations from normal (Table 1).

The elemental compositions of the grains were determined with the PHI 700 Auger Nanoprobe. The areas of interest were initially sputter-cleaned by scanning a 2 kV, 1 μm diameter $Ar^+$ ion beam over a broad area (~2 mm) to remove atmospheric surface contamination. Auger electron energy spectra, covering the energy range of 30–1730 eV, were obtained with a 10 kV 0.25 nA primary electron beam. The compositions of the grains were calculated from the peak-to-peak heights of the differentiated spectra using elemental sensitivity factors determined from olivine and pyroxene standards (Stadermann et al. 2009). We also acquired Auger elemental distribution maps (e.g., O, C, Si, Fe, S) of the grains.

Raman spectroscopy was used to obtain mineralogical information on grain LAP-149. We used the Renishaw inVia© Raman imaging system, with a laser wavelength of 532 nm for excitation that was condensed by a long working distance 100× objective (NA = 0.8) to <1 μm spot of 2.5 mW at the sample. The spectra were acquired in a spectral window of 100 – 4000 $\Delta cm^{-1}$. While transmission electron microscopy is traditionally used to study the internal structure of presolar graphites, recent studies have demonstrated that Raman spectroscopy can also be used to obtain direct *in situ* structural information about presolar carbonaceous grains (e.g., graphites, SiC), without additional sample preparation (Wopenka et al. 2013). The Raman spectra of carbonaceous grains exhibit two first-order main bands: the D-band (D for disordered) at 1332 $cm^{-1}$ and the G-band (G for graphite) at 1582 $cm^{-1}$. The ratio of the relative heights of the D- and G-bands provides direct information on the degree of structural order. Raman analysis of more than a hundred presolar graphites from three different density fractions (2.15-2.20 $g/cm^3$) in the Murchison meteorite indicates that only about 50% of presolar graphite grains have spectra consistent with well-crystallized $sp^2$-bonded carbon graphite (D/G < 0.5) while about 30% of the grains have spectra consistent with disordered graphite (D/G = 0.5-1.1). The remaining grains can be classified into three groups: glassy carbon (D/G > 1.1), unusual $sp^2$-bonded graphitic carbon and non-crystalline kerogen-like carbon (Wopenka et al. 2013).

The CO nova models discussed below were computed with SHIVA, a one-dimensional implicit hydrodynamical code in Lagrangian formulation, extensively applied to the modeling of stellar explosions (e.g., classical novae, X-ray bursts, Type Ia supernovae). The code solves the standard set of differential equations of stellar evolution: conservation of mass, momentum, and energy; energy transport by radiation and convection; and the definition of the Lagrangian velocity. A time-dependent formalism for convective transport has been included whenever the characteristic convective timescale becomes larger than the integration timestep. Partial mixing between adjacent convective shells is handled by a diffusion equation. The equation of state includes contributions from the (degenerate) electron gas, the ion plasma, and radiation. Coulomb corrections to the electronic pressure are also taken into account. Radiative and conductive opacity are considered. The code is linked to a reaction network that contained about 120 nuclear species, ranging from H to $^{48}$Ti through 630 nuclear reactions, with updated rates. See José (2016) and José & Hernanz (1998) for additional details.

3. RESULTS

Grain LAP-141 is relatively small with a diameter of about 300 nm and is characterized by an enrichment in $^{12}$C relative to the solar isotopic ratio; its oxygen isotopic composition is solar within 2σ errors (Figure 1, Table 1). It also exhibits depletions in $^{33,34}$S ($\delta^{33}$S = -107 ± 23, $\delta^{34}$S = -130 ± 11, Table 1). This grain has an Auger elemental spectrum consistent with graphite (Figure 2). Auger elemental maps show that the small Fe and S peaks in the spectrum reflect the presence of a tiny (~80 nm in diameter) iron sulfide subgrain inside LAP-141. Quantification indicates that the subgrain has a pyrrhotite-like composition (Fe = 76 ± 8 at.% and S = 24 ± 2 at.%). It is likely that the anomalous S isotopic composition of LAP-141 mostly reflects the composition of the iron sulfide subgrain.

Grain LAP-149 is larger, with a diameter of about 1 μm, and has one of the lowest $^{12}$C/$^{13}$C ratios (1.41 ± 0.01) ever measured; it also has a high $^{14}$N/$^{15}$N ratio (941 ± 81; Figure 1). Such a low $^{12}$C/$^{13}$C ratio (less than the CNO cycle nuclear equilibrium value of 3.5) is extremely rare among presolar grains. Only one other graphite grain, KFC1b-202, has a similarly low $^{12}$C/$^{13}$C ratio; however, its N and O isotopic compositions are consistent with solar values (Figure 1; Amari et al. 2014). The oxygen and silicon isotopic compositions of LAP-149 ($^{16}$O/$^{17}$O = 2594 ± 228, $^{16}$O/$^{18}$O = 516 ± 19, $\delta^{29}$Si = -8 ± 24, $\delta^{30}$Si = -23 ± 29) are also solar within uncertainties. The Auger elemental spectrum of LAP-149 is also consistent with graphite (Figure 2). In addition, its Raman spectrum is different from the carbonaceous material present in the surrounding matrix but resembles the spectra of well-crystallized presolar graphite grains, with a D/G ratio of ~0.51 (Figure 3). Perfectly stacked sp$^2$-bonded graphitic carbon sheets exhibit only the G-band peak; the presence of a D-band peak indicates either structural defects in the grain or damage from the NanoSIMS Cs beam (Wopenka et al. 2013).

## 4. DISCUSSION

**4.1 Origin of LAP-141 and its iron sulfide subgrain: Low-metallicity AGB star or Type II supernova?**

While previous studies have suggested the possible existence of presolar FeS grains (Heck et al. 2012; Hoppe et al. 2012; Hynes et al. 2011), to the best of our knowledge, no such grains have been unambiguously identified. We report here the first confirmed identification of a presolar iron sulfide subgrain inside a small graphite grain (LAP-141; ~300 nm in diameter).

Based on their $^{12}C/^{13}C$ ratios, Amari et al. (2014) divided presolar graphite grains from Murchison into three distinct populations. LAP-141 is characterized by a high $^{12}C/^{13}C$ ratio (537 ± 24) and belongs to Population III ($^{12}C/^{13}C > 200$). Presolar graphites from this population can condense in the envelopes of low-metallicity asymptotic giant branch (AGB) stars or in the ejecta of Type II (core collapse) supernovae (Amari et al. 2014; Jadhav et al. 2013a; Lodders & Amari 2005).

Asymptotic giant branch (AGB) stars are a major source of presolar graphite grains (~50%) (Jadhav et al. 2014). Comparison of the high $^{12}C/^{13}C$ ratio, excess $^{32}S$ (negative $\delta^{33,34}S$ values) and solar O isotopic composition of LAP-141 with Torino AGB star models (Figure 4), suggests that LAP-141 could have condensed in the stellar atmosphere of a low-metallicity AGB star (Amari et al. 2014; Jadhav et al. 2013a; Xu et al. 2015). Indeed, the C, O and S isotopic compositions of LAP-141 can all be reproduced by nucleosynthesis models of a 3.0 M$_\odot$ AGB star with Z = 0.003 or 0.006 (Figure 4). If this is the case, the S isotopic composition of LAP-141 would require a parent star that was characterized by negative $\delta^{33,34}S$ values relative to solar.

However, an origin in a Type II supernova is also possible. The isotopic compositions of LAP-141 (and its Fe-sulfide subgrain) are clearly similar to supernova SiC grains (SiC X and C grains, Figure 4b,c), which are both believed to have condensed in the ejecta of Type II supernovae. Pre-supernova massive stars have an onion-like layered structure, with a Ni-rich zone, a Si/S zone, oxygen-rich zones (O/Si, O/Ne, and O/C), and He/C and He/N zones, surrounded by a H envelope (Meyer et al. 1995). A large range of sulfur isotopic compositions is observed in the different layers. Like presolar SiC X grains, the S isotopic composition of LAP-141 could be the signature of the Si/S zone (Hoppe et al. 2012; Xu et al. 2015). However, this zone is not favorable for the condensation of carbonaceous grains (C/O << 1) and, thus, mixing of different supernova zones is required to explain the condensation of both the graphite grain and its Fe-sulfide subgrain. Because the Fe-sulfide subgrain is contained inside the graphite grain, it must have condensed before its host graphite; it is thus possible that the two grains formed at different places within the supernova ejecta. This is consistent with recent observations of large-scale transport and asymmetric mixing within the Cassiopeia A supernova ejecta (Grefenstette et al. 2014). The negative $\delta^{33,34}$S values observed in grain LAP-141 might also reflect the production of $^{32}$S by the decay of the short-lived $^{32}$Si ($\tau_{1/2}$ = 153 years) in the O/C or He/C supernova regions (Fujiya et al. 2013; Pignatari et al. 2013; Xu et al. 2015). This process was proposed to explain the large $^{32}$S excesses observed in presolar SiC C grains (Hoppe et al. 2015; Pignatari et al. 2013).

Calculations of grain condensation in circumstellar envelopes and stellar ejecta predict the formation of iron sulfide (Lodders & Fegley, 1999), which is consistent with the observation of a pyrrhotite-like iron sulfide subgrain in LAP-141. Indeed, equilibrium condensation calculations suggest the formation of significant amounts of FeS in the circumstellar envelopes of red giant stars under carbon-rich conditions (C/O > 1) (Lodders, 2006) and models of dust condensation in core-collapse supernova ejecta also predict the presence of FeS grains (Sarangi & Cherchneff 2015). In particular, Cherchneff & Dwek (2010) suggested that significant amounts of FeS can condense in the inner Ni/Si/S zone of unmixed supernova ejecta; this model is consistent with the predictions of negative $\delta^{33,34}$S values, similar to those of LAP-141, in the Si/S zone (Xu et al. 2015).

Astronomical observations have also suggested the presence of FeS in circumstellar envelopes and stellar ejecta. FeS has been proposed as a possible carrier for several features in the IR spectra of carbon-rich AGB stars (Hony et al. 2003, 2002). Spectral fitting of the observed 21 μm-peak dust feature in the infrared (IR) spectrograph spectra of the Cassiopeia A Supernova Remnant acquired by the NASA *Spitzer Space Telescope* also indicates the possible presence of FeS grains (e.g., Rho et al. 2008).

Ultimately, while we cannot exclude an origin in a low-metallicity AGB star, the very close similarity of the carbon and sulfur isotopic compositions of LAP-141 (and its Fe-sulfide subgrain) with presolar SiC X and C grains lead us to favor an origin in the ejecta of a Type II supernova.

## 4.2 LAP-149: Extremely $^{13}$C-rich presolar graphite from a low-mass CO nova ejecta

The origin of LAP-149 is restricted to several types of stars that can produce very low $^{12}$C/$^{13}$C ratios: these include born-again AGB stars, J-type stars, novae and core-collapse (Type II) supernovae. Born-again AGB stars are late-stage stars, which experienced a very late thermal pulse (VLTP). During the VLTP, mixing of leftover hydrogen from the convective envelope down to the hot helium intershell of the star induces H-burning, which converts $^{12}$C into $^{13}$C, reducing the $^{12}$C/$^{13}$C ratio (Herwig et al. 2011). However, the minimum $^{12}$C/$^{13}$C ratios predicted by VLTP theoretical models are higher than that of LAP-149 and, even if they were lower, the model trends are inconsistent with the high $^{14}$N/$^{15}$N observed in LAP-149 (Jadhav et al. 2013b);

in addition, neutron capture associated with the VLTP is expected to produce large excesses in $^{30}$Si (Liu et al. 2014) that are not observed in LAP-149.

We also considered formation of LAP-149 in a Type II supernova because $^{12}$C/$^{13}$C ratios as low as 2 are predicted in some zones of a supernova. However, model calculations that reproduce the $^{12}$C/$^{13}$C ratio of LAP-149 are typically characterized by C/O ratios much below unity, precluding the condensation of C-rich grains (Amari et al. 2014; Jadhav et al. 2013a; Rauscher et al. 2002). New supernova explosive He shell models yield both very low $^{12}$C/$^{13}$C ratios and C/O > 1 in the He/C zone (Pignatari et al. 2015), but these models also predict extremely low $^{14}$N/$^{15}$N ratios, the opposite of what is seen in LAP-149. Our grain also does not show other typical supernova isotopic signatures such as excesses in $^{18}$O and $^{28}$Si (Jadhav et al. 2013a).

Astronomical observations have shown the presence of a small group of carbon stars, called J-type stars, which are characterized by strong $^{13}$CN and $^{13}$C absorption bands with low $^{12}$C/$^{13}$C ratios (3-15). While their origins are still enigmatic, spectral studies show that they have solar s-process elemental abundances and are characterized by a large range of $^{14}$N/$^{15}$N ratios (Figure 1; Hedrosa et al. 2013). Theoretical models suggest that the low $^{12}$C/$^{13}$C ratios of low-mass J stars (M ~2-3 M$_\odot$) are caused by thermal pulses that mix material from the stellar envelope down to the H-burning shell where $^{13}$C is produced in the CNO cycle ("cool bottom processing") (Nollett et al. 2003). However, cool bottom processing can only decrease the $^{12}$C/$^{13}$C ratio of the envelope down to the nuclear equilibrium (steady-state) value of ~3.5 for the CNO cycle. Lower $^{12}$C/$^{13}$C ratios can be obtained in higher-mass stars (M ≥ 4 M$_\odot$) in which the higher temperatures and pressures allow hot hydrogen burning to take place at the bottom of the convective envelope (called "hot bottom processing"). While this process decreases the $^{12}$C/$^{13}$C ratio of the envelope, the destruction of $^{12}$C also reduces the C/O ratio, leading to O-rich conditions that, again, preclude the condensation of carbonaceous grains like LAP-149.

Finally, several studies have suggested that presolar grains with very low $^{12}$C/$^{13}$C ($\leq 5$) ratios may originate in nova ejecta (Amari et al. 2001; Clayton & Hoyle, 1976; Heck et al. 2009; José & Hernanz, 2007; José et al. 2004). A nova explosion occurs when a white dwarf (WD) core accretes enough material from a nearby companion star to cause rapid fusion of the accreted hydrogen and trigger an explosion (José, 2016). The very low $^{12}$C/$^{13}$C ratios (0.3-3) observed in novae are due to the $^{12}$C(p, $\gamma$)$^{13}$N($\beta^+$)$^{13}$C chain reaction during burning of the accreted hydrogen (José & Hernanz, 2007). These predictions of very low $^{12}$C/$^{13}$C ratios are consistent with limits on the $^{12}$C/$^{13}$C ratio estimated from fitting of $^{13}$CO and $^{12}$CO bands in the infrared spectra of several nova ejecta (Banerjee & Ashok, 2012; Banerjee et al. 2016; Evans & Rawlings, 2008). Based on the composition of the WD core, we distinguish two types of novae: CO novae, for initial masses of the primary star below ~8 M$_\odot$, which undergo hydrogen and helium burning, leaving a carbon- and oxygen-rich WD core; and ONe novae for slightly more massive stars (8-10 M$_\odot$) which, in addition, undergo carbon burning, leaving an oxygen- and neon-rich WD core (José, 2016; José et al. 2004). Stellar nucleosynthesis models of ONe novae cannot reproduce the isotopic compositions of grain LAP-149: ONe novae ejecta are characterized by very low $^{12}$C/$^{13}$C ratios (0.73-1.1), consistent with the $^{12}$C/$^{13}$C ratio of LAP-149, but also have very low $^{14}$N/$^{15}$N (0.25-3.6) ratios (José et al. 2004), in contrast with the $^{14}$N-rich isotopic composition of this grain.

However, CO nova models appear to be more promising (José et al. 2004). We computed seven new one-dimensional CO nova models with WD masses ranging between 0.6-1.15 M$_\odot$ and mixing fractions of 25% or 50% between the outer layers of the WD core and material accreted from the companion star. While most models predict isotopic compositions that are not consistent with those of LAP-149 (e.g., low $^{14}$N/$^{15}$N), one model with a WD mass of 0.6 M$_\odot$ and a mixing fraction of 50% predicts C and N isotopic compositions ($^{13}C/^{12}C = 2$ and $^{14}N/^{15}N = 979$) that are virtually identical to those of LAP-149 (Figure 1). This model is also consistent with the solar silicon and sulfur isotopic compositions of LAP-149 because peak temperatures in CO novae are not high enough to significantly modify the Si and S isotopic compositions (José et al. 2004; Lodders & Amari, 2005). Thus, LAP-149 represents the first plausible grain of CO nova origin and its C, N, Si, and S isotopic compositions can be reproduced by pure nova ejecta, without any extra mixing with solar composition material (Figure 1). This differs from most nova grains candidates, which require large amounts of solar composition material (> 90%) to be mixed with the nova ejecta in order to reproduce their isotopic compositions (even for major elements such as carbon) (Amari et al. 2001; Nittler & Hoppe, 2005).

Oxygen represents a unique problem. All nova models predict extreme oxygen isotopic compositions with very large excesses in $^{17}O$ and depletions in $^{18}O$. For example, the CO nova model consistent with the C and N isotopic compositions observed in LAP-149 predicts a $^{16}O/^{17}O$ ratio of 352 and a $^{16}O/^{18}O$ ratio of 416,300, clearly inconsistent with the solar values observed in LAP-149. Such extreme O isotopic anomalies are also not observed in any other grains that have been suggested to originate from novae, including O-rich presolar grains (Gyngard et al. 2010b; Leitner et al. 2012). Previous studies have argued that the close-to-normal nitrogen, oxygen and silicon isotopic compositions of presolar graphite grains reflect isotopic equilibration by either chemical processing in the laboratory, or secondary aqueous/thermal alteration on their meteorite parent body asteroids (Groopman et al. 2012; Stadermann et al. 2005). However, LAP-149 was found in situ and did not undergo the chemical isolation procedures experienced by graphite grains from the Murchison and Orgueil meteorites. In addition, it is important to keep in mind that Murchison and Orgueil, the sources for all other presolar graphite grains, are well-known to have been highly affected by aqueous alteration (e.g., Le Guillou et al. 2014), with the common presence of secondary minerals, such as phyllosilicates and carbonates, identified in their matrix. In contrast, LAP 031117 has experienced very limited thermal and/or aqueous alteration (Chizmadia & Cabret-Lebron, 2009); secondary minerals (e.g., phyllosilicates and carbonates) are extremely rare and were not observed in the area where the two graphite grains were identified and, thus, extensive isotopic equilibration is unlikely for LAP-149.

It has previously been shown that during NanoSIMS analysis of presolar grains identified in thin sections and densely packed grain dispersions, for small presolar grains (200-300 nm in diameter), the partial overlap of the primary beam onto both the grain and isotopically normal surrounding material can significantly dilute the measured isotopic compositions of the grain, by shifting its measured isotopic ratio toward the normal solar value (Nguyen et al. 2003, 2007). However, grain LAP-149 is ~1 um in diameter and Nguyen et al. (2007) show that this dilution effect is limited for grains larger than about 600 nm (with a shift in $\delta^{18}O$ of less than 15 ‰). As shown in Figure 5, the profile of the O isotopic ratios across the grain and the surrounding matrix area show that there is no O isotopic gradient in the grain, suggesting that the contribution of oxygen from the surrounding material is insignificant in LAP-149. Moreover, we were conservative in the integration area chosen for the reported isotopic ratios (Table 1) and considered only the central core region of the grain to be certain to avoid any contribution from isotopically normal surrounding material (Figure 5). Re-deposition of oxygen from surrounding areas onto the surface of the grain by the sputtering process is also unlikely as no variation of the O isotopic compositions is observed between the different layers. Finally, we also identified several small presolar silicate grains (250-300 nm in diameter), less than 50-100 μm away from the two graphite grains, which exhibit large $^{17}O$ excesses (Haenecour et al. 2015). Even if O is only a trace element in graphite and a major element in silicate grains, it seems unlikely that one grain would get completely isotopically homogenized without homogenizing other neighboring grains (and without leaving any evidence of alteration in the surrounding matrix).

The discrepancies between the models and the grain data for O isotopic compositions remain a puzzling characteristic for all nova grains, and underscore the need for additional modeling to understand the O isotopic composition of these grains. Unlike previous studies, the fact that we can rule out isotopic equilibration of our grain by secondary alteration or laboratory treatment is an important result that future studies will need to consider in nova models.

Finally, we consider elemental constraints. Astronomical observations indicate the presence of dust around most CO novae and spectral fitting suggests the presence of silicates, SiC, carbon and hydrocarbons (Gehrz et al. 1998). The 0.6 $M_\odot$ CO nova model consistent with the C and N isotopic compositions of LAP-149 predicts a C/O ratio above unity (C/O = 1.1), favorable for the condensation of carbonaceous grains, such as graphite. While condensation likely takes place dynamically in a nova environment, we computed equilibrium condensation sequences to provide preliminary estimates on the type of condensates and their condensation temperatures in low-mass CO nova ejecta for the relevant range of total pressures ($10^{-5} - 10^{-8}$ bars) expected for the ejecta (José et al. 2004; Lodders, 2003; Lodders & Amari, 2005; Lodders & Fegley, 1993). The major abundance differences for a low-mass CO nova, relative to solar, are the large increases in C (~200 times), N (~150 times) and O (90 times). Cooling of such a gas with an enhanced atomic ratio of C/O = 1.1 mostly leads to C-bearing condensates (José et al. 2004; Lodders & Fegley, 1995). Our calculations indicate that graphite is predicted to condense from the gas under these conditions, and is stable above 1900K for all total pressures considered here (Figure 6). Moreover, while the exact condensation sequence of other phases depends on pressure, graphite is always the only expected condensate for about 900K below its initial condensation temperature, before SiC, AlN, $Al_2O_3$, and CaS condense (Figure 6). At $10^{-7}$ bar, the sequence is: C (2040K), SiC (1191K), AlN (1156K), CaS (1089K), and $Al_2O_3$ (1061K). At $10^{-5}$ bar, it is slightly different: C (2229K), AlN (1289K), CaS (1217K), $Al_2O_3$ (1204K), and SiC (1182 K). The large thermal stability range of graphite observed in our models demonstrates that presolar graphite grains are expected to condense and remain stable in low-mass CO nova ejecta. Taken together, the evidence strongly suggests that LAP-149 condensed in the ejecta of a low mass CO nova, and represents the first plausible CO nova grain identified in the presolar grain population.

**4.3 Presolar Graphite Abundance in Type 3.0 CO Chondrite**

Presolar graphite grains are among the least abundant presolar grain type, with bulk meteoritic abundances of about 1 ppm (Amari et al. 2014), although large variations (0.08 – 13.1 ppm, Figure 7) are observed between different carbonaceous and ordinary chondrite groups (Huss et al. 2003). There is no clear correlation between these abundance variations and the meteorite's petrographic type. Indeed, the abundance of presolar graphite in Murchison is significantly higher (between ~1 and 4.7 ppm, Amari et al. 2014; Huss et al. 2003) than the one in ALHA77307 (CO3.00), although Murchison has experienced much more aqueous alteration. The reason for this is unclear, but might reflect heterogeneous distribution of these grains in their meteorite parent bodies.

Based on the sizes of the two presolar graphites identified, the total fine-grained area mapped (67,000 $\mu m^2$) in LAP 031117 and an average matrix abundance in CO3 chondrites of about 30% (Krot et al. 2014), we estimate a bulk presolar graphite abundance of $5^{+7}_{-3}$ ppm (Figure 7); the errors on this estimate are 1σ and are based on counting statistics (Gehrels, 1986). This is the first presolar graphite abundance estimate based on direct *in situ* ion imaging.

While our estimate is associated with large uncertainties, it is at least an order of magnitude higher than the abundance of 0.08 ppm determined for ALHA77307, which was calculated from the abundances of noble gases in a meteorite acid residue (Huss et al. 2003). It is unlikely that this difference is due to different degrees of secondary processing (thermal or aqueous alteration) between LAP 031117 and ALHA77307, because both meteorites are CO3.00 chondrites that are only minimally affected by secondary processing and have similar abundances of both presolar silicates (~160 ppm) and SiC (~30 ppm) (Haenecour et al. 2015). Similar discrepancies between abundance estimates based on NanoSIMS raster ion imaging and noble gas measurements have also been reported for presolar SiC grains (Davidson et al. 2014). Possible explanations for the discrepancies are a dependence of noble gas abundances on grain size, partial loss of noble gases from some of the grains, or a heterogeneous distribution of presolar grains in fine-grained matrix areas (Davidson et al. 2014).

## 5. SUMMARY AND CONCLUSION

Our work on the *in situ* coordinated isotopic, elemental and microstructural analyses of two graphite grains in the CO3.00 chondrite LAP 031117 can be summarized as follows:

- The isotopic composition of grain LAP-149 (extremely $^{13}$C-rich and $^{15}$N-poor) is in excellent agreement with the predictions from a 0.6 $M_\odot$ CO nova model. LAP-149 thus represents the first plausible grain of CO nova origin, confirming that dust from novae contributed to the original building blocks of the solar system. The O isotope discrepancies between the stellar nucleosynthesis models and all nova candidates remain puzzling, and the fact that, unlike previous studies, we can rule out isotopic equilibration and/or mixing with solar composition material for our grain is an important result that future studies will need to consider in nova models.

- We report the first confirmed identification of a presolar iron sulfide subgrain inside a small graphite grain. Its carbon and sulfur isotopic compositions are similar to SiC X and C grains, suggesting an origin in the ejecta of a type II supernova.

- Based on the two graphite grains identified in LAP 031117, we estimate a bulk presolar graphite abundance of $5^{+7}_{-3}$ ppm in this meteorite, consistent with the abundance in Murchison but significantly higher than an estimate based on noble gases in the ALHA77307 meteorite.

**Acknowledgements:** This work was funded by NASA Earth and Space Science Fellowship NNX12AN77H (P.H.), NASA Grants NNX14AG25G (C.F.), NNX16AD31G (S.A.), NNX13AM22G (A.W.) and NNX13AH08G (F.G.), MINECO grants AYA2013-42762-P and AYA2014-59084-P and AGAUR (J.J.), and NSF Grant 1517541 (K.L.). We thank T. Smolar of Washington University for maintenance of the NanoSIMS 50 and Auger Nanoprobe. We are also grateful for a thorough and helpful review of the manuscript by an anonymous referee.

# TABLE

**Table 1**
Isotopic compositions of graphites and predictions from CO nova models*

| | Name | $^{12}C/^{13}C$ | $^{14}N/^{15}N$ | $^{16}O/^{17}O$ | $^{16}O/^{18}O$ | $\delta^{29}Si$ | $\delta^{30}Si$ | $\delta^{33}S$ | $\delta^{34}S$ | |
|---|---|---|---|---|---|---|---|---|---|---|
| | LAP-141 | 537 ± 24 | - | 2861 ± 147 | 520 ± 11 | - | - | -107 ± 23 | -130 ± 11 | - |
| | LAP-149 | 1.41 ± 0.01 | 941 ± 81 | 2594 ± 228 | 516 ± 19 | -8 ± 24 | -23 ± 29 | -23 ± 143 | 6 ± 70 | - |
| | KFC1b-202 | 2.1 ± 0.1 | 273 ± 53 | - | 556 ± 65 | - | - | - | - | - |
| | $M_\odot$    F | | | | | | | | | C/O |
| CO nova models* | 0.6    0.5 | 2 | 979 | 352 | 416300 | 0 | 7 | -0.57 | -0.13 | 1.1 |
| | 0.8    0.25 | 0.4 | 147 | 120 | 121154 | -5 | 3 | - | - | 0.5 |
| | 0.8    0.5 | 0.6 | 88 | 132 | 170122 | -5 | 3 | - | - | 0.9 |
| | 1    0.25 | 0.5 | 10 | 41 | 77180 | -57 | 14 | - | - | 0.7 |
| | 1    0.5 | 0.5 | 12 | 52 | 100746 | -34 | 17 | - | - | 0.8 |
| | 1.15    0.25 | 0.9 | 2 | 12 | 30063 | -482 | 46 | - | - | 0.8 |
| | 1.15    0.5 | 0.7 | 4 | 21 | 55750 | -230 | 116 | - | - | 0.7 |
| | Standard value | 89 | 272 | 2625 | 499 | - | - | - | - | - |

*C/O ratios calculated from the mass fractions of $^{12,13}C$ and $^{16,17,18}O$ in each CO nova model; $M_\odot$ = solar masses; F = mixing fraction; see text for details. Data for KFC1b-202 from Amari et al. (2014). $\delta^{29,30}Si$ and $\delta^{33,34}S$ are expressed in permil (‰) where $\delta_{sample} = \left(\frac{R_{sample} - R_{standard}}{R_{standard}}\right) \times 1000$. Errors are 1σ.

**FIGURES CAPTIONS**

**Figure 1.** (a) Carbon and oxygen ($^{12}C/^{13}C$ and $^{18}O/^{16}O$) and (b) carbon and nitrogen isotopic compositions ($^{12}C/^{13}C$ and $^{14}N/^{15}N$) of LAP-141 and LAP-149 compared with other presolar graphites (Amari et al. 2001, 2014; Groopman et al. 2012; Hoppe et al. 1995; Jadhav et al. 2013b; Nicolussi et al. 1998). Also shown are results from CO nova models (see text for details), predictions from ONe nova models (José et al. 2004), and astronomical observations of J-type stars (Hedrosa et al. 2013). The nitrogen isotopic composition of LAP-141 was not measured.

**Figure 2.** Differentiated Auger spectra and elemental distribution maps showing the C-rich natures of LAP-141 and LAP-149. LAP-141 exhibits small Fe and S peaks due to the presence of an iron sulfide subgrain. Also shown is the secondary electron (SE) image for LAP-149 and a false-color image of the $^{12}C/^{13}C$ ratio measured in the NanoSIMS.

**Figure 3.** Raman spectrum of grain LAP-149 compared with surrounding matrix material in LAP 031117, a presolar graphite grain from Murchison (Wopenka et al. 2013), and a graphite standard from the RRUFF database (Lafuente et al. 2015).

**Figure 4.** (a) Carbon and oxygen isotopic compositions of LAP-141 compared with two Torino AGB models (Bisterzo et al. 2010; Gallino et al. 1998): 3.0 $M_\odot$ AGB with Z = 0.003 (red line) and Z = 0.006 (green line). Also shown are presolar graphite grains from previous studies (Amari et al. 2001, 2014; Groopman et al. 2012; Hoppe et al. 1995; Jadhav et al. 2013a; Nicolussi et al. 1998). (b, c) Carbon ($^{12}C/^{13}C$) and sulfur ($\delta^{33, 34}S$, expressed in ‰) isotopic composition of LAP-141 compared with presolar SiC grains from the literature: Fujiya et al. (2013); Gyngard et al. (2006, 2010a, 2012); Hoppe et al. (2012); Orthous-Daunay et al. (2012); Xu et al. (2015) and the two Torino AGB models. Error bars for LAP-141 are 2σ.

Figure 5. Profiles across grain LAP-149 and the surrounding matrix area of the NanoSIMS total oxygen-to-carbon count signal ratio ($^{16,17,18}O^-/^{12,13}C^-$), the estimated Auger elemental O/C ratio, and the $^{17}O/^{16}O$ and $^{18}O/^{16}O$ ratios. The Auger O/C ratio for point 3 is directly derived from the Auger spectrum, as the integration area for the Auger spectrum overlaps with point 3, and the other five points were extrapolated by comparison with the NanoSIMS $O^-/C^-$ ratios.

**Figure 6**. Initial equilibrium temperature (K) and condensation sequences showing the different types of grains expected to form in the ejecta of a 0.6 $M_\odot$ CO nova for total pressures of $10^{-5}$ and $10^{-7}$ bars.

**Figure 7.** Presolar graphite abundances in LAP 031117 compared with abundance estimates for several carbonaceous and ordinary chondrites (Amari et al. 2014; Huss & Lewis, 1995; Huss et al. 2003).

**Figure 1**

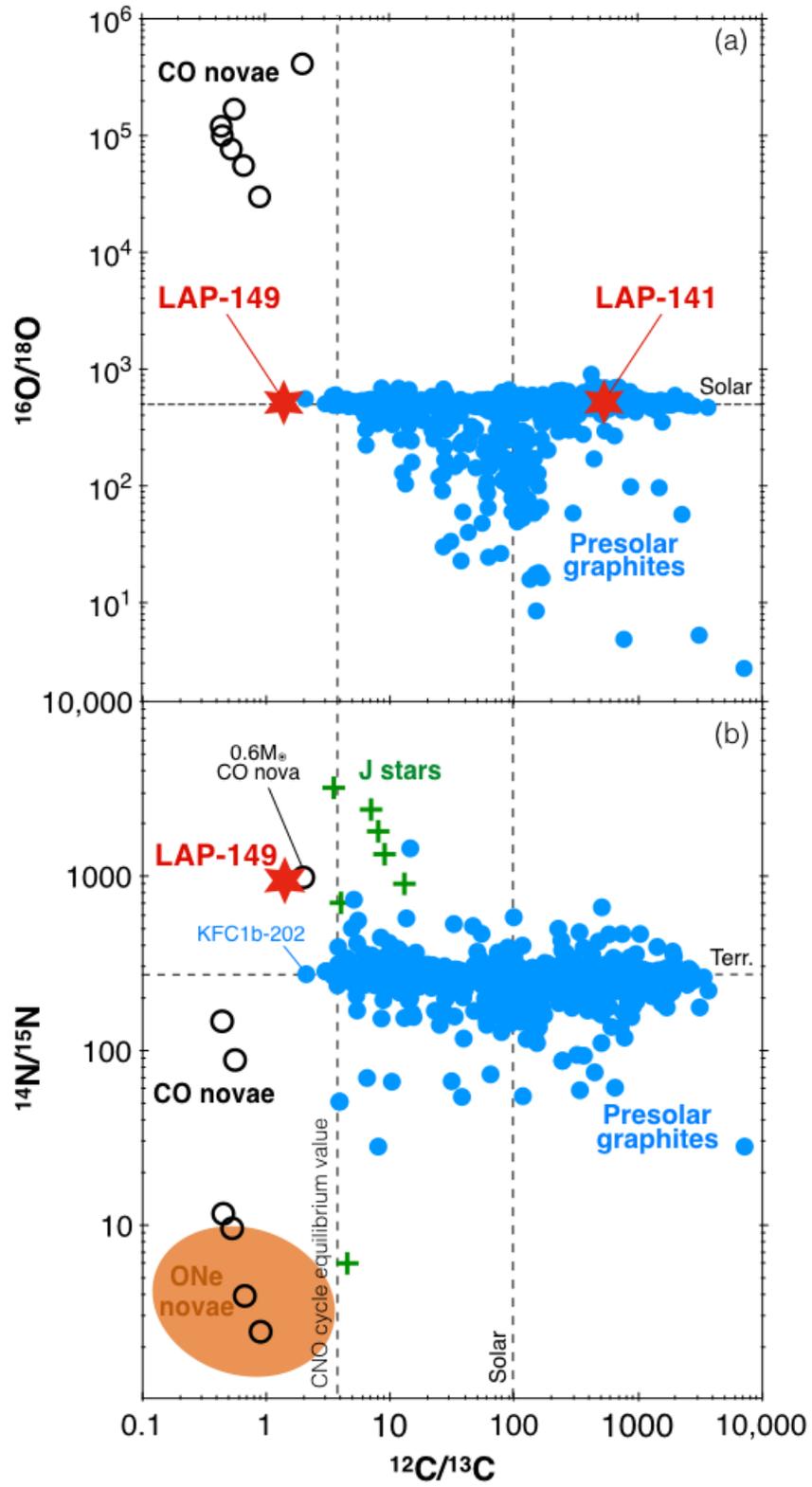

**Figure 2**

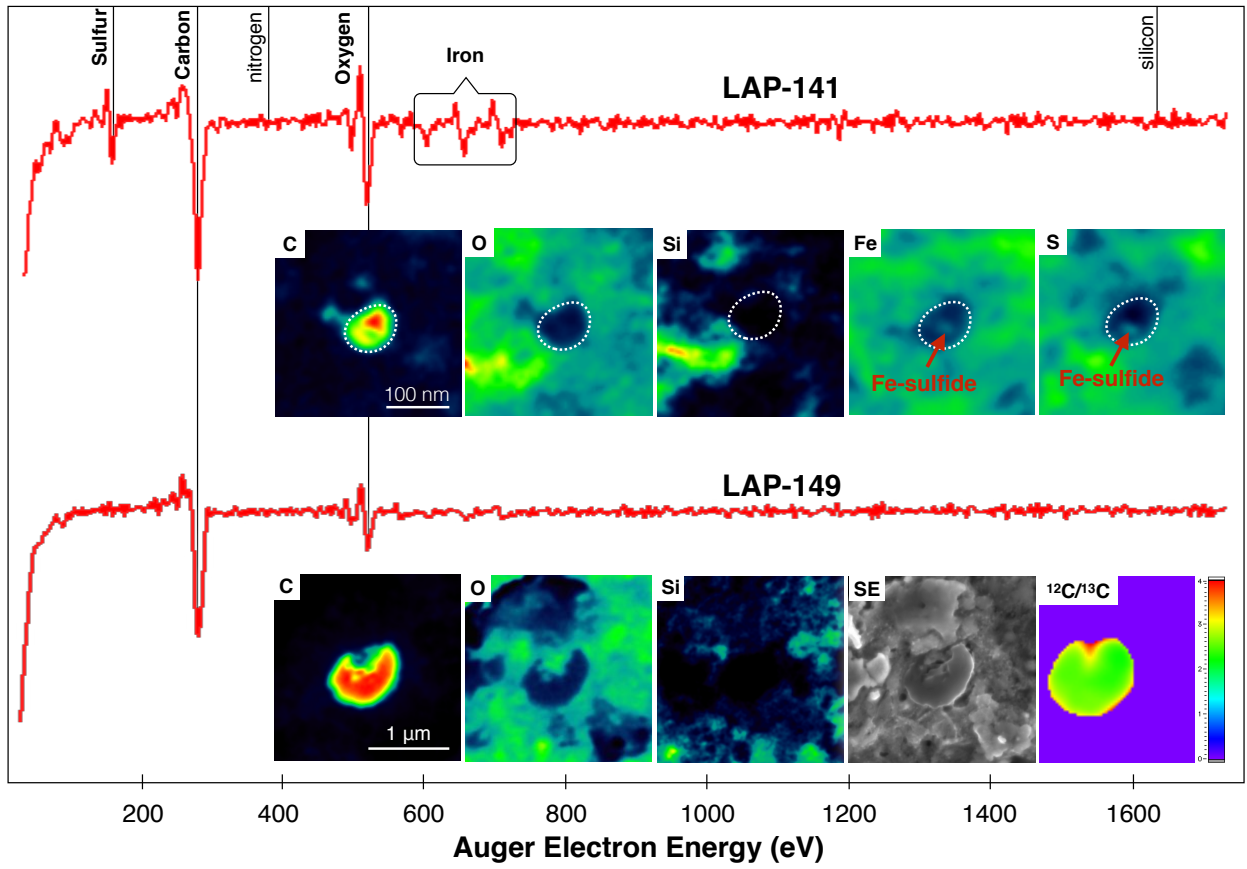

**Figure 3**

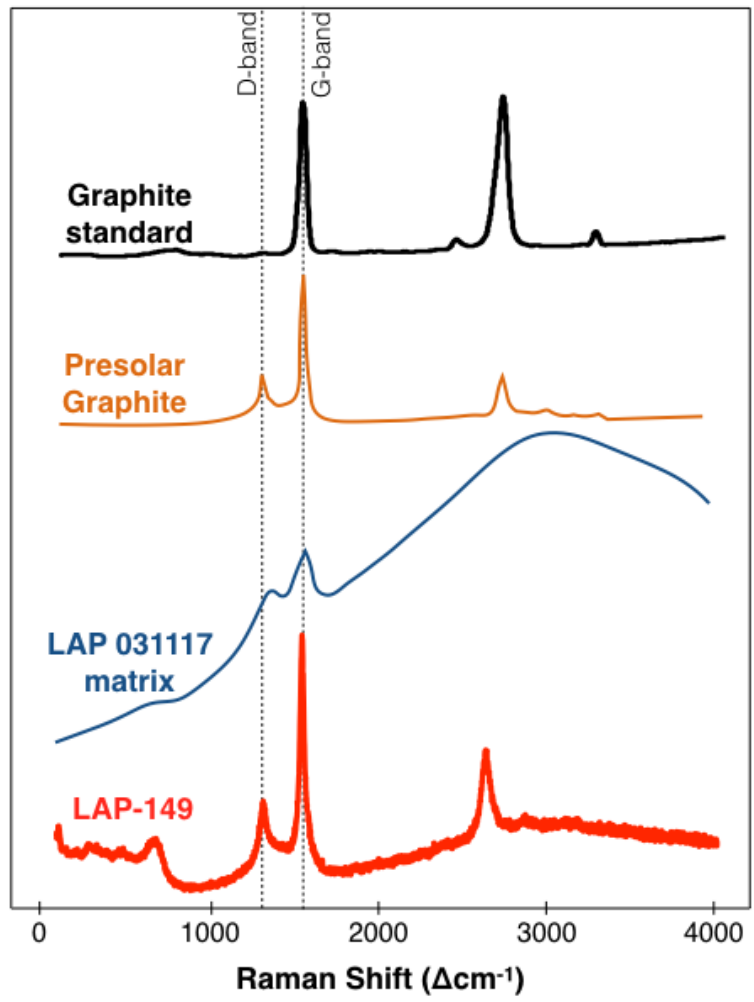

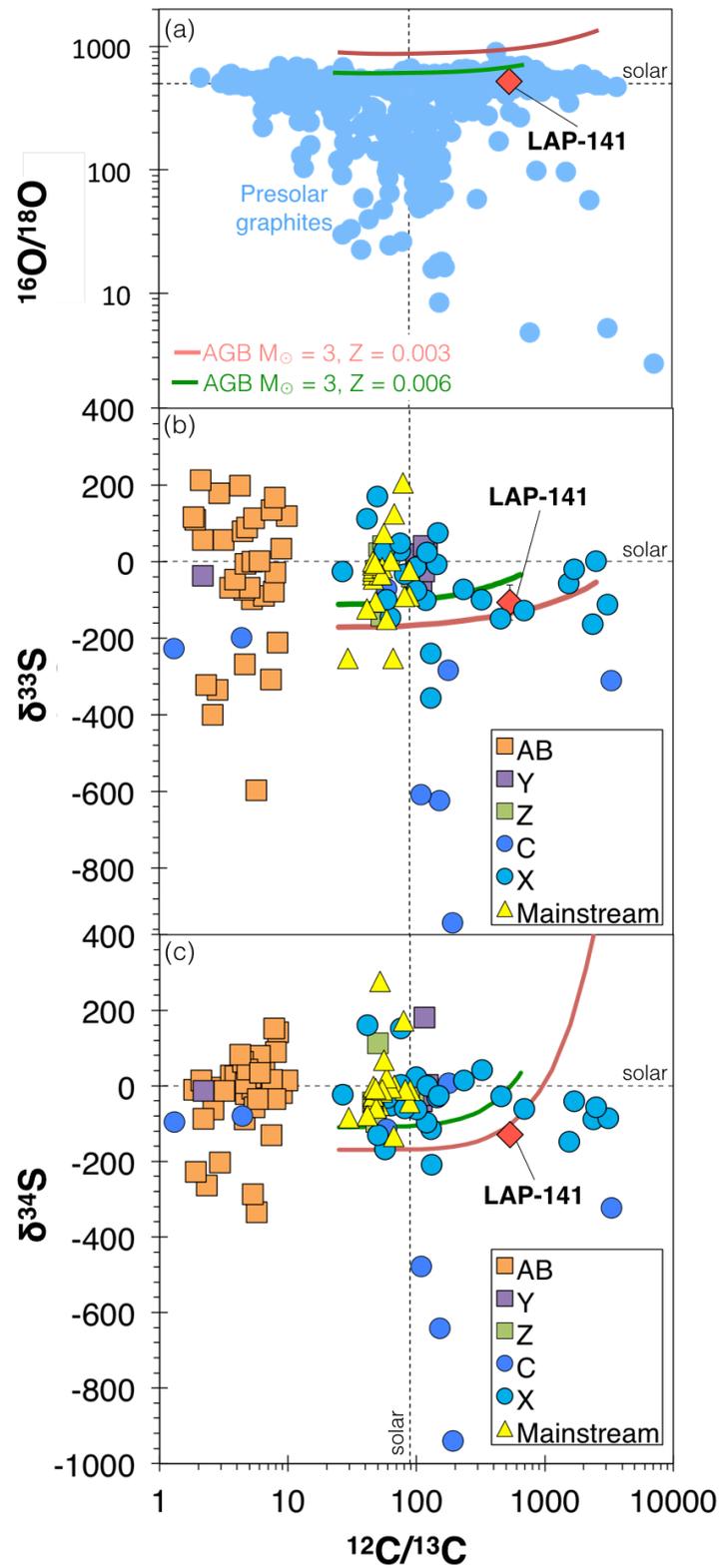

**Figure 5**

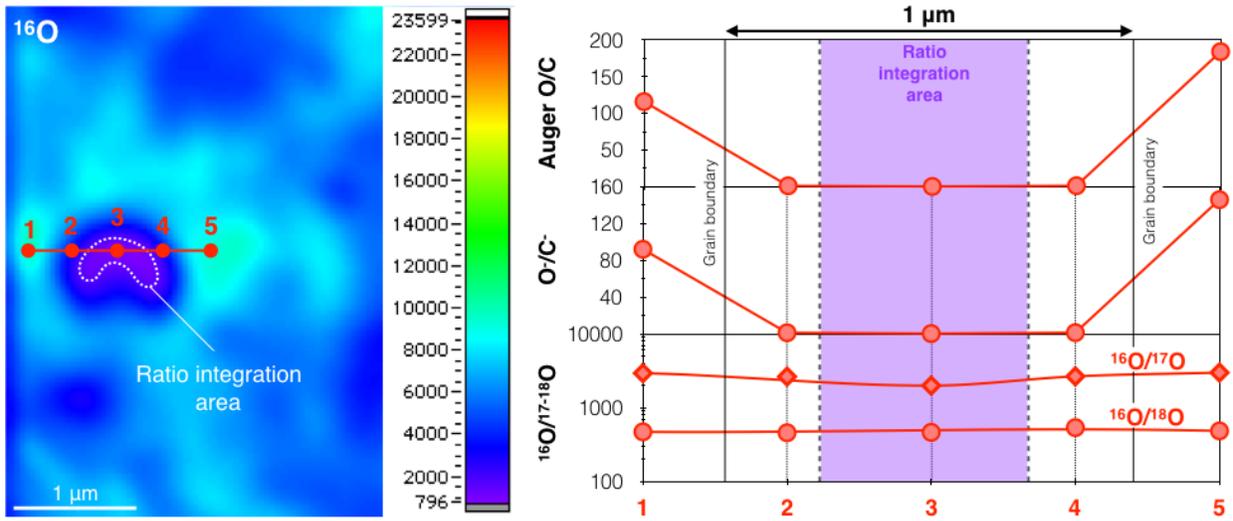

**Figure 6**

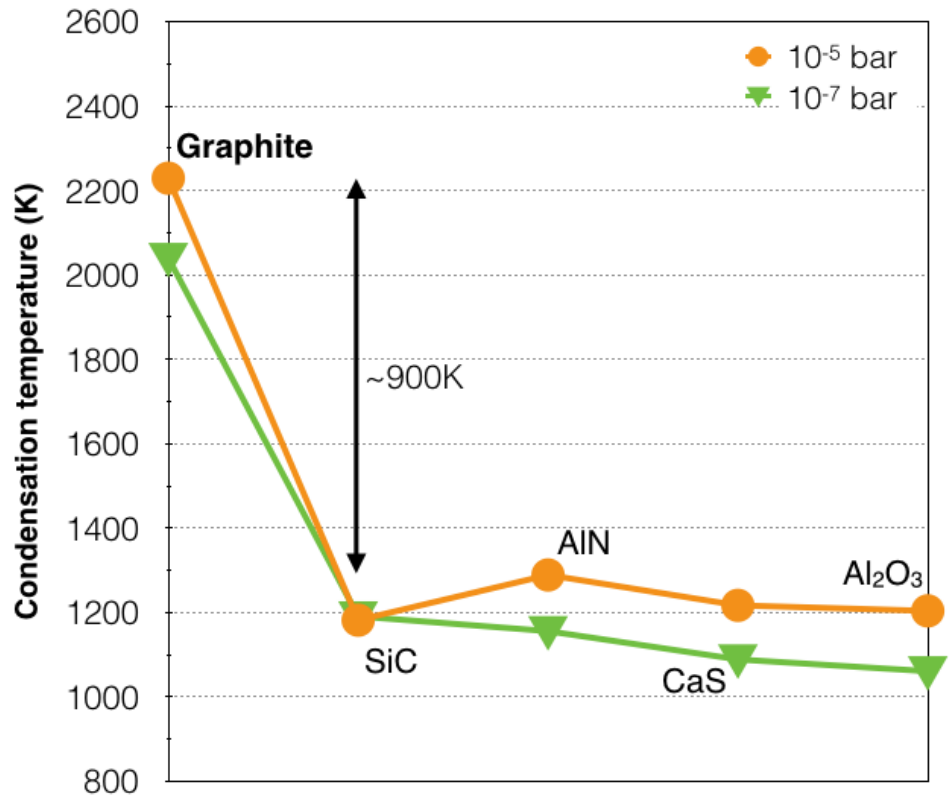

**Figure 7**

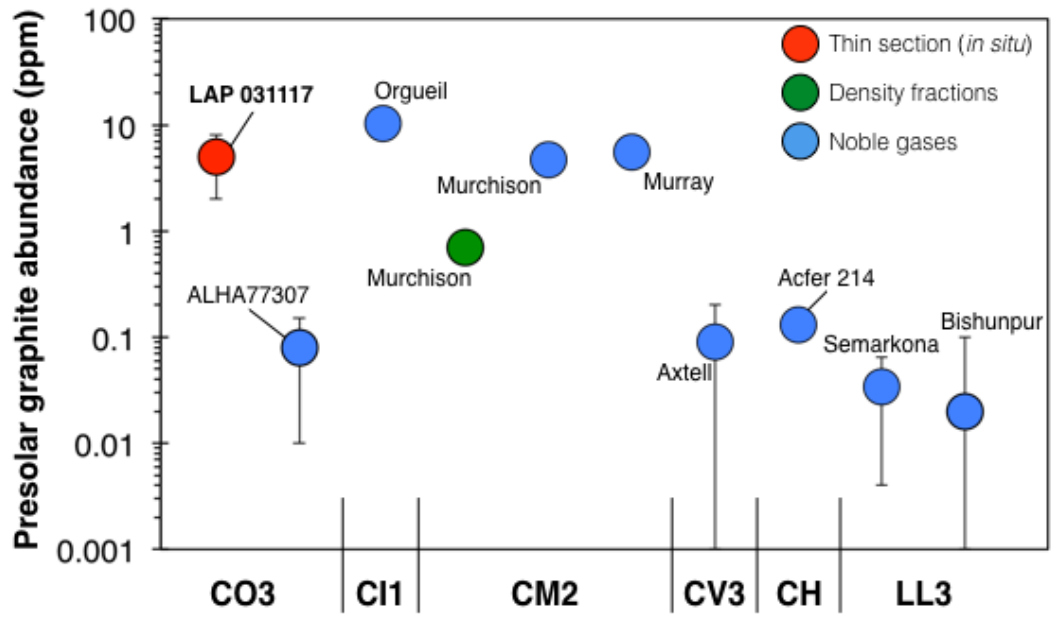